 \documentclass[preprint,floats,aps,epsfig,nofootinbib,amssymb]{revtex4}

\usepackage{slashed}
\usepackage{slashed}
\usepackage{graphicx,color}
\usepackage{epsfig}
\usepackage{subfigure}
\usepackage{epsfig}
\usepackage{dcolumn}
\usepackage{bm}
\usepackage{color}

\newcommand{\beq}{\begin {equation}}
\newcommand{\eeq}{\end   {equation}}
\newcommand{\bea}{\begin {eqnarray}}
\newcommand{\eea}{\end   {eqnarray}}
\newcommand{\baa}{\begin {array}   }
\newcommand{\eaa}{\end   {array}   }
\newcommand{\bit}{\begin {itemize} }
\newcommand{\eit}{\end   {itemize} }
\newcommand{\be }{\begin {equation}}
\newcommand{\ee }{\end   {equation}}

\begin{document}

\preprint{ACFI-T15-21}
\preprint{IPMU15-0213}



\title{The Minimal Scalar-Stealth Top Interpretation of the Diphoton Excess}

\author{Wei Chao$^a$}
\email{chao@physics.umass.edu}
\author{Ran Huo$^b$}
\email{ran.huo@ipmu.jp}
\author{Jiang-Hao Yu$^a$}
\email{jhyu@physics.umass.edu}

\affiliation{$^a$ Amherst Center for Fundamental Interactions, Department of Physics, University of Massachusetts-Amherst, Amherst, MA 01003, USA}
\affiliation{$^b$  Kavli IPMU (WPI), UTIAS, The University of Tokyo, Kashiwa, Chiba 277-8583, Japan}

\begin{abstract}
	
A resonance near $750$ GeV has been observed at both the ATLAS and CMS experiments in its diphoton channel. We propose an interpretation based on a simplified model, in which a $750$ GeV scalar singlet and minimal stealth vector-like top are introduced. At the LHC, the new scalar is produced through gluon fusion, and subsequently decays to diphoton final state, with the new vector-like quarks running in the loop diagram. The lightness of the stealth top could enhance the diphoton rate significantly, while still keeping other channels under the current LHC search limits. We find that the stealth quarks with their masses around $400 \sim 500$ GeV and couplings around $0.5 \sim 1.5$ could explain the diphoton excess, while avoid current di-jet constraints. We also comment on other potential signatures of this stealth model.

\end{abstract}

\maketitle
\section{Introduction}

Recently both the ATLAS and CMS experiments observed diphoton channel excesses with the LHC run-II data at $\sqrt{s}=13$ TeV~\cite{ATLAS13,CMS13}. The ATLAS experiment~\cite{ATLAS13} detected 14 events peaked at $747$ GeV from an integrated luminosity of 3.2 fb$^{-1}$, which corresponds to a $3.6\sigma$ local statistical significance. The CMS experiment~\cite{CMS13} on the other hand detected 10 events peaked at $760$ GeV from an integrated luminosity of 2.6 fb$^{-1}$, which corresponds to a $2.6\sigma$ local statistical significance. Roughly they are consistent with their respective previous $\sqrt{s}=8$ TeV measurements~\cite{ATLAS8,CMS8}, in which the ATLAS experiment~\cite{ATLAS8} did not go beyond an invariant mass of $600$ GeV, and the CMS experiment~\cite{CMS8} observed an $\sim1.8\sigma$ local significance there.

A $\sigma(pp\to S)\text{Br}(S\to\gamma\gamma)\sim5-10$ fb is found to be consistent with combination of both experiments. On the other hand, the diphoton excess is accompanied neither by missing transverse energy or leptons or jets, nor by other channels at the same invariant mass such as the $ZZ$, $WW$, $\ell^+ \ell^-$ or $jj$ channels. This could be an argument against the reality of the diphoton excess, meaning nothing but a statistical fluctuations; but if taken seriously, it put strong constraint on the implied model building~\cite{Angelescu:2015uiz,Backovic:2015fnp,Bellazzini:2015nxw,Buttazzo:2015txu,DiChiara:2015vdm,Ellis:2015oso,Franceschini:2015kwy,Gupta:2015zzs,Harigaya:2015ezk,Higaki:2015jag,Knapen:2015dap,Low:2015qep,Mambrini:2015wyu,McDermott:2015sck,Molinaro:2015cwg,Nakai:2015ptz,Petersson:2015mkr,Pilaftsis:2015ycr}~\footnote{
Refs.~\cite{Cao:2015pto, Matsuzaki:2015che, Kobakhidze:2015ldh, Cox:2015ckc, Ahmed:2015uqt, Agrawal:2015dbf, Martinez:2015kmn, Becirevic:2015fmu, No:2015bsn, Demidov:2015zqn, Fichet:2015vvy, Bian:2015kjt,Curtin:2015jcv, Chakrabortty:2015hff, Csaki:2015vek,  Falkowski:2015swt, Aloni:2015mxa, Bai:2015nbs} on this subject appeared at the same time as this work.}.

According to the Landau-Yang theorem~\cite{Landau:1948kw,Yang:1950rg} the $750$ GeV diphoton resonance could only correspond to a spin 0 or 2 particle. Here we will focus on the spin 0 case, with further assumption that it is CP even.
The diphoton signal could only be induced by a loop diagram, with an electromagnetically charged particle running in it.
However even a nonzero coupling of the scalar to the particles running in the diphoton loop is there, the tree level two body decay resonance is not seen in any other channels, that could minimally be interpreted as the particle running in the loop are too heavy to be produced on-shell.
This implies a lower bound of $375$ GeV to the mass  of the particle running in the loop.
Regarding to the production mechanism, the minimal choice is the new scalar with $750$ GeV mass is singly and directly produced via the gluon fusion. 
As will be seen in section IV, compared to the quark initial state, indeed the gluon gluon initial state is favored by a parton distribution function (PDF) rescaling.
Furthermore, the new scalar should not have tree-level coupling to the Standard Model (SM) light quarks unless it is extremely suppressed,
otherwise it will have both tree-level decay to the light quarks which is not favored by the PDF rescaling, and will presumably be seen as a dijet resonance.
Then the simplest mechanism is that the loop of the decay to diphoton is the same as the loop of the production  in the gluon fusion.
If we really identify the two to be the same, it could be a heavy top as a simple fermionic choice, which must be vector-like (VL); or a scalar top as a bosonic choice.
In this paper taking the fermionic choice, we reach our minimal model content to explain the diphoton excess: the singlet scalar and VL top, as proposed in Ref.~\cite{Xiao:2014kba, Xiao:2015tja}.

Still opening options are the VL top representation under the SM $SU(2)_L$, here we consider two cases: an $SU(2)_L$ singlet VL top alone, and an $SU(2)_L$ singlet plus doublet VL tops. Numerically in both cases we find that to fit a $\sigma(pp\to S)\text{Br}(S\to\gamma\gamma)$ of $\sim5-10$ fb, a combination with the current general VL top direct search bound of about $700-900$ GeV will lead to a too large Yukawa coupling, out of the perturbative regime.
This could be cured by either adding new degree of freedom, or opening some low mass region for the VL top.
We take the second choice, in which the VL top in the hidden sector carries an odd-$Z_2$ charge.
This odd VL top is motivated in the little Higgs model with $T$-parity~\cite{Cheng:2003ju,Cheng:2004yc}, top flavored dark matter~\cite{Kilic:2015vka}, vector fermion-portal dark matter model~\cite{Yu:2014pra}, etc.
To lower the VL top mass, we assume that the VL top is nearly degenerate with a dark matter candidate.  We refer to this VL top as the {\it stealth  top}, similar to stealth stop~\cite{Fan:2011yu}. This setup makes it evading the general $700-900$ GeV bound~\cite{VLconstraint} for the VL quark.
The light stealth  top could enhance the diphoton rate significantly, while still keeping other channels under the current LHC search limits.
We find that the stealth tops around $400-500$ GeV with its coupling around $0.5-1.5$ could explain the diphoton excess, and satisfy all the other constraints such as from the dijet channel.
If the dominant decay channels only are loop-induced diphoton and dijet final states, the
total width of the new scalar will be very narrow.

The paper is organized as following. In Section \ref{model} we present the model in some detail. Next in Section \ref{EWPT} we discuss some general constraints of the model such as LHC, LUX, and LEP constraints. We present our main fitting work to the $\sigma(pp\to S)\text{Br}(S\to\gamma\gamma)$ of $\sim5-10$ fb in Section \ref{Diphoton}, followed by a check in other channels in Section \ref{OtherChannel}. At last we discuss and conclude our paper.

\section{The Model}\label{model}

In our simplest setup, the SM is extended to incorporate a singlet scalar $S$ and a VL top quark $T$~\cite{Xiao:2014kba, Xiao:2015tja}. The VL top quark $T$ is motivated in various models such as the little Higgs~\cite{ArkaniHamed:2002qy}, composite Higgs~\cite{Agashe:2004rs}, and extra dimensions~\cite{Randall:1999ee}, etc. The scalar $S$ is identified as the $750$ GeV resonance. It is a singlet under the SM gauge group $SU(2)_L \times U(1)_Y$.
The scalar $S$ couples to the heavy top quark~\cite{Xiao:2014kba, Xiao:2015tja} via the Yukawa coupling
\bea
	{\mathcal L}_{\rm Yuk} \supset y_T S \bar{T}_L T_R + h.c., \quad \quad ({\textrm{\bf Model I}})
\eea
Similar to the Higgs production, the new scalar is produced via the VL top quark loop.

We assign the VL top quark $T$ a $Z_2$-odd charge, and keep the SM particles and $S$ to be $Z_2$-even.
This $Z_2$-odd VL top quark is known in little Higgs model with $T$-parity~\cite{Cheng:2003ju,Cheng:2004yc}, top flavored dark matter~\cite{Kilic:2015vka}, vector fermion-portal dark matter model~\cite{Yu:2014pra}, etc.
Similar to the littlest Higgs model with $T$-parity, we introduce a color neutral scalar $X$ or vector $X_\mu$ as the dark matter candidate.
If the dark matter does not interact with the new scalar, or $m_X > m_S/2$, so that two body decay is kinematically forbidden, the dark matter is irrelevant to our study of purpose.
We will focus on stealth VL top quark, similar to the stealth stop quark, in which
the  VL top quark $T$ and the dark matter are almost degenerate.
In this case, the VL top quark can be as light as several hundred GeVs.
In the following we denote this VL top quark as the {\it stealth top}.

Comparing  to the usual VL quark, the stealth quark can significantly enhance production cross section of the scalar $S$ in the gluon fusion process, as shown in the section~\ref{EWPT}.
We know that in the gluon fusion process, the production cross section is quite sensitive to the mass of the particle running in the loop.
The searches on the VL quark at the LHC push its mass up to $700-900$ GeV~\cite{VLconstraint}.
However, the stealth quark could be very light if the VL top quark $T$ and the dark matter are almost degenerate.
This can be seen from the LHC searches on the $T$-odd quark~\cite{Aad:2012uu}.

To further enhance the production cross section of the new scalar, we could introduce additional doublet $Q$ as the partner of the left-handed quark doublet.
This is quite similar to the left-handed stop quark doublet in the supersymmetry.
We refer to this fermion setup as our model II.
Its relevant Lagrangian is
\bea
	{\mathcal L}_{\rm Yuk} \supset y_T S \bar{T}_L T_R + y_Q S \bar{Q}_L Q_R + h.c., \quad \quad ({\textrm{\bf Model II}})
\eea
For simplicity, in this work we take $y_Q = y_T$, and the masses $m_Q = m_T$.
We also want to focus on the case that the doublet $Q$ is stealth, due to their degenerate masses with the dark matter.
Both the stealth singlet $T = (T_L, T_R)$ and the stealth doublet $Q = (Q_L, Q_R)$ are free of the $U(1)_Y$ anomaly because they are VL fermions.

The new scalar $S$ can only interact with the SM through its mixing with the SM Higgs boson.
We assume the mixing is quite small by setting the coupling in the term
\bea\label{mixing}
{\mathcal L} \supset \lambda_{HS} S^2 H^\dagger H,
\eea
small after spontaneous symmetry breaking.
Thus we expect the (tree level) branching ratios of the $S \to WW/ZZ$ and $S \to hh$ is small, as discussed in Section~\ref{OtherChannel}.

\section{Model Constraints}\label{EWPT}

\begin{figure}
  \includegraphics[width=0.4\textwidth]{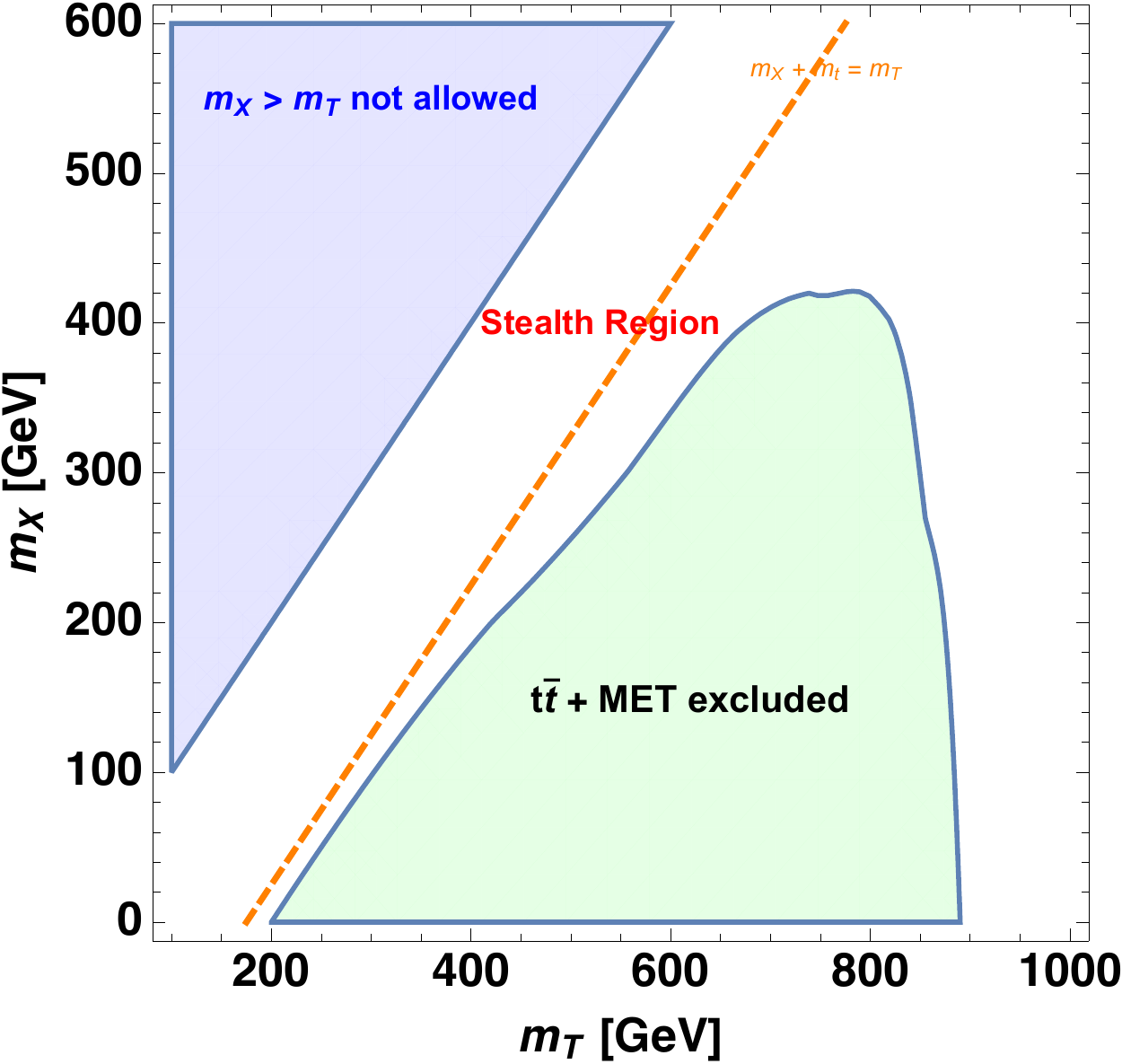}
  \includegraphics[width=0.4\textwidth]{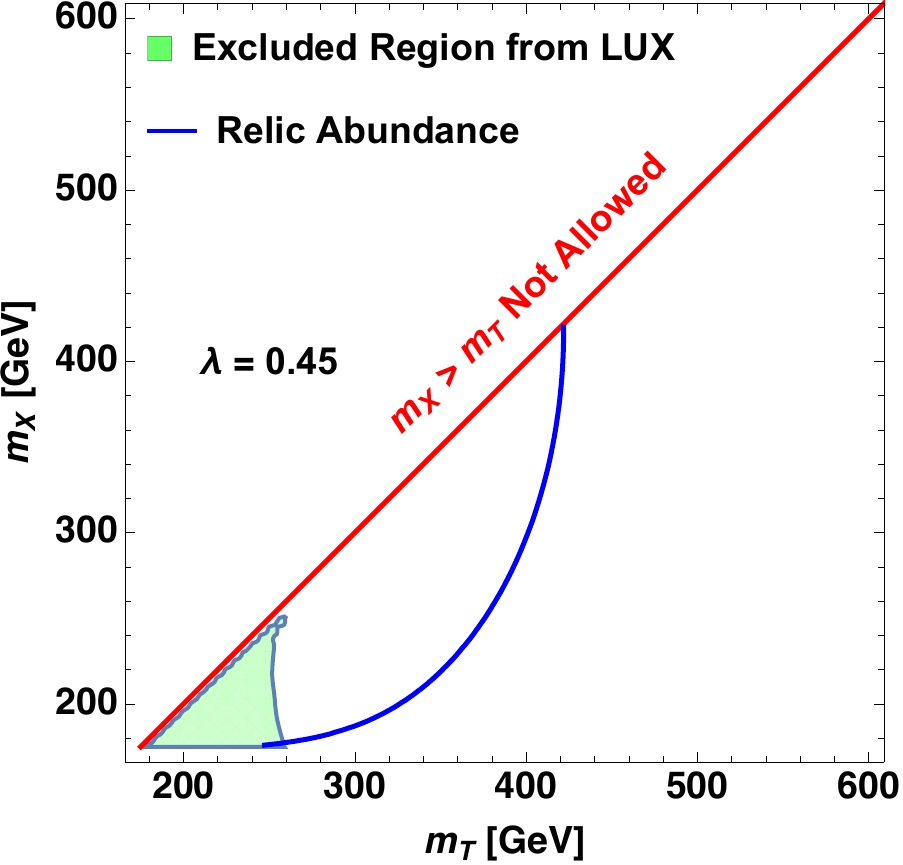}
\caption{\label{lhcbound} We show the parameter contour ($m_T, m_X$) and its exclusion from the collider searches (left) and dark matter searches (right). On the left, the exclusion limit on the heavy top $T$ are shown in the green region from the ATLAS searches on top pair plus missing energy final state~\cite{Aad:2014bva}. The blue region is  kinematically forbidden. The region where the heavy top and the dark matter $X$ are degenerate is still allowed and marked as "stealth region".  On the right, the exclusion from the LUX data~\cite{Akerib:2013tjd} are shown in green colored region. The blue curve is that when the coupling between the dark matter and the heavy top is 0.5, the required mass region to obtain the correct relic density. }
\end{figure}

The $Z_2$-odd VL top have been searched by the ATLAS~\cite{Aad:2012uu} experiment, but the search is based on 7 TeV data and the limit is quite loose.
Note that the stop searches have the same final state: the top quark pair plus missing energy, 
so we expect that the stop search limit could put strong bound on the VL top quark.
%
Here we take the exclusion limit from  the ATLAS~\cite{Aad:2014bva} experiment and convert it into the exclusion on $Z_2$-odd VL top, as shown in Fig.~\ref{lhcbound} (left).
Since the VL top could have two-body and three-body decay, the stealth region is the mass range with the following mass difference between the VL top and the dark matter
\bea
	\Delta m = m_T - m_X \lesssim 180\,\, {\textrm GeV}.
	\label{eq:stealthmass}
\eea
From the Fig.~\ref{lhcbound} (left) We see the stealth region is very hard to be excluded, so we mark the degenerated region as the stealth region.
Thus the stealth top evades the direct searches from the current LHC searches.

In the UV completed model, one may worry about the constraints from the dark matter relic density and direct detection searches.
One UV completion of our model is the top flavored dark matter~\cite{Kilic:2015vka} with the Lagrangian
\bea
	{\mathcal L}_{\rm DM} \supset \lambda \bar{T}_L X t_R + h.c.
\eea
where the dark matter $X$ is a scalar.
In this specific UV setup, we calculate the relic density and direct detection bounds, as shown in Fig.~\ref{lhcbound} (right).
In our calculation, we refer to the latest direct detection limit from LUX~\cite{Akerib:2013tjd}.
%
We find that the parameter space in the stealth region could still satisfy the relic density and fullfil the current LUX limit.
Due to the loop suppression in the direct detection calculation, the direct detection limit is quite weak.
We refer reader to Ref.~\cite{Kilic:2015vka} for detailed UV model and discussions.


For the EW precision measurements constraint as well as other SM Higgs digluon and diphoton constraints, in the effective field theory approach the contribution from the VL top are all proportional to coupling $y'^2_T$ for Peskin-Takeuchi $S$ and Higgs digluon and diphoton deviation, or $y'^4_T$ for the Peskin-Takeuchi $T$~\cite{Huo:2015exa}, where $y'_T$ is the Yukawa coupling to the SM Higgs. Since we are not specifying $y'_T$ they can be arbitrarily small to avoid constraints. There is also no correction when integrating out our hypercharge zero singlet scalar $S$~\cite{Chiang:2015ura}. For model II which has nontrivial $SU(2)_L$ doublet, the only precision measurement constraint is the tri-gauge coupling, which can be universally calculated by~\cite{Henning:2014wua} and generate a Wilson coefficient of $\frac{1}{6}g\epsilon^{abc}W^{a\mu}_\rho W^{b\nu}_\mu W^{c\rho}_\nu$, but the constraints should be weak.

\section{Fitting the Diphoton Excess}\label{Diphoton}

In this section our goal is to fit the signal cross section at $\sqrt{s}=13$ TeV LHC
\bea
\sigma(pp\to S)\text{Br}(S\to\gamma\gamma)\sim5-10 ~\text{fb},
\eea
while still respecting the previous $\sqrt{s}=8$ TeV LHC constraint $\sigma(pp\to S)\text{Br}(S\to\gamma\gamma)<1.32$ fb. We first consider one VL top case, then extend to the two VL top case~\footnote{For the dark matter phenomenology induced by these VL tops, which is interesting but out of the reach of this paper, we refer the reader to Ref~\cite{Kilic:2015vka}.}. As can be seen from the section II, the VL top has Yukawa interaction with $S$, which induces the effective couplings of $SG_{\mu\nu}^a G_{a}^{\mu\nu}$ and $SF_{\mu\nu} F^{\mu \nu}$ at one loop level. The effective coupling can be written as
\begin{eqnarray}
C_{sgg} &=&{g_s^2 \over 16 \pi^2 } {y_T \over m_T } A_{1/2} \left( \tau_T \right) \\
C_{s\gamma\gamma } &=& { e^2 \over 16 \pi^2 } \left[  \sum 2 n_C Q^2   {y_T \over m_T } A_{1/2} \left( \tau_{T}\right)\right],
\end{eqnarray}
where~\cite{Hunter} $\tau_T\equiv4m_T^2/m_S^2$ and
\bea
A_{1/2}(\tau)=-2\tau(1+(1-\tau)f(\tau)),\qquad\qquad f(\tau)=\left\{\begin{array}{ll}
\left(\sin^{-1}\sqrt{\frac{1}{\tau}}\right)^2, & \tau\geq1\\
-\frac{1}{4}\left(\log\frac{1+\sqrt{1-\tau}}{1-\sqrt{1-\tau}}-i\pi\right)^2, & \tau<1
\end{array}\right..
\eea

The decay partial widths are given by
\begin{eqnarray}
\Gamma_{\rm gg} &=&  {\alpha_s^2 m_S^3 \over 128 \pi^3 }  \left|{y_T \over m_T } A_{1/2} \left( \tau_T \right) \right|^2,\\
\Gamma_{\rm\gamma\gamma} &=& {\alpha^2 m_S^3 \over 1024 \pi^3 }  \left |  \sum 2 n_C Q^2   {y_T \over m_T } A_{1/2} \left( \tau_{T}\right)\right |^2.
\end{eqnarray}
The diphoton width are much smaller, since we have the same form factor $A_{1/2}(\tau_T)$ for the two amplitudes, the ratio of the two can be calculated straightforwardly from the ratios of the QCD strong coupling to the QED electromagnetic coupling as well as the group factor to electric charge, which is
\bea\label{ggtogaga}
\frac{\Gamma_{\rm gg}}{\Gamma_{\rm\gamma\gamma}}=148.8.
\eea

Note that in the minimal setup the total width could be roughly given by the digluon width $\Gamma_S\simeq\Gamma_{\rm gg}$. The diphoton amplitude are always accompanied by the $ZZ$ coupling and $Z\gamma$ coupling, but they are at most the same order in branching ratio with $\gamma\gamma$, and the current ATLAS and CMS bounds are either much weaker than that of the diphoton channel or not performed, here we will ignore them. Due to the Eq.~(\ref{mixing}) there is in principle tree level decay of $S$ into a pair of SM Higgses, and the mixing induces every 750 GeV SM Higgs decay channel such as the $\bar{b}b$, $WW$ and $t\bar t$, but they are all proportional to the mixing parameter $\lambda_{HS}^2$ and thus can be set to be very small. The mixing will provide freedom to fit the total decay width, but we will see that increasing the total width will make the $\text{Br}(S\to\gamma\gamma)$ even smaller. Thus it is more difficult to fit the model parameter, and here we simply ignore them. In Section \ref{OtherChannel} we will go back to this point. Since we have no other couplings of $S$ (except for our dark matter candidate $X$ corresponding to invisible decay width), that enumerates the decay channel.

As for production, the cross section of a $750~{\rm GeV}$ SM-like Higgs boson is~\cite{Heinemeyer:2013tqa}
\begin{eqnarray}
&&\sigma^{\rm SM-like} (gg\to S) @{\rm 8 ~TeV}  = 156.8~{\rm fb}\; , \\
&&\sigma^{\rm SM-like} (gg\to S) @{\rm 13 ~TeV}  = 742.76~{\rm fb}\; .
\end{eqnarray}
The ratio between the two is just a gluon PDF rescaling, which is $R_{gg}=4.7$ and can fit the ratio of the aforementioned $\sqrt{s}=13$ TeV signal to the $\sqrt{s}=8$ TeV constraint. On the other hand, a similar light quark PDF rescaling gives $R_{qq}=2.5$, which is inconsistent with the ratio. This justifies our choice of gluon fusion production against direct quark pair production.

Since the Higgs production via gluon-gluon fusion is proportional to the decay rate of the Higgs to gluon gluon, one can scale the cross section of the new scalar to the SM-like Higgs, which can be written as
\begin{eqnarray}
\sigma(gg\to S) = {\Gamma_{S\to gg} \over \Gamma_{S\to gg}^{\rm SM}} \sigma^{\rm SM} (gg\to S).
\end{eqnarray}
Given this formula, eventually one can check the parameter regime of this model as an explanation of the LHC diphoton excess.

\begin{figure}
  \includegraphics[width=0.45\textwidth]{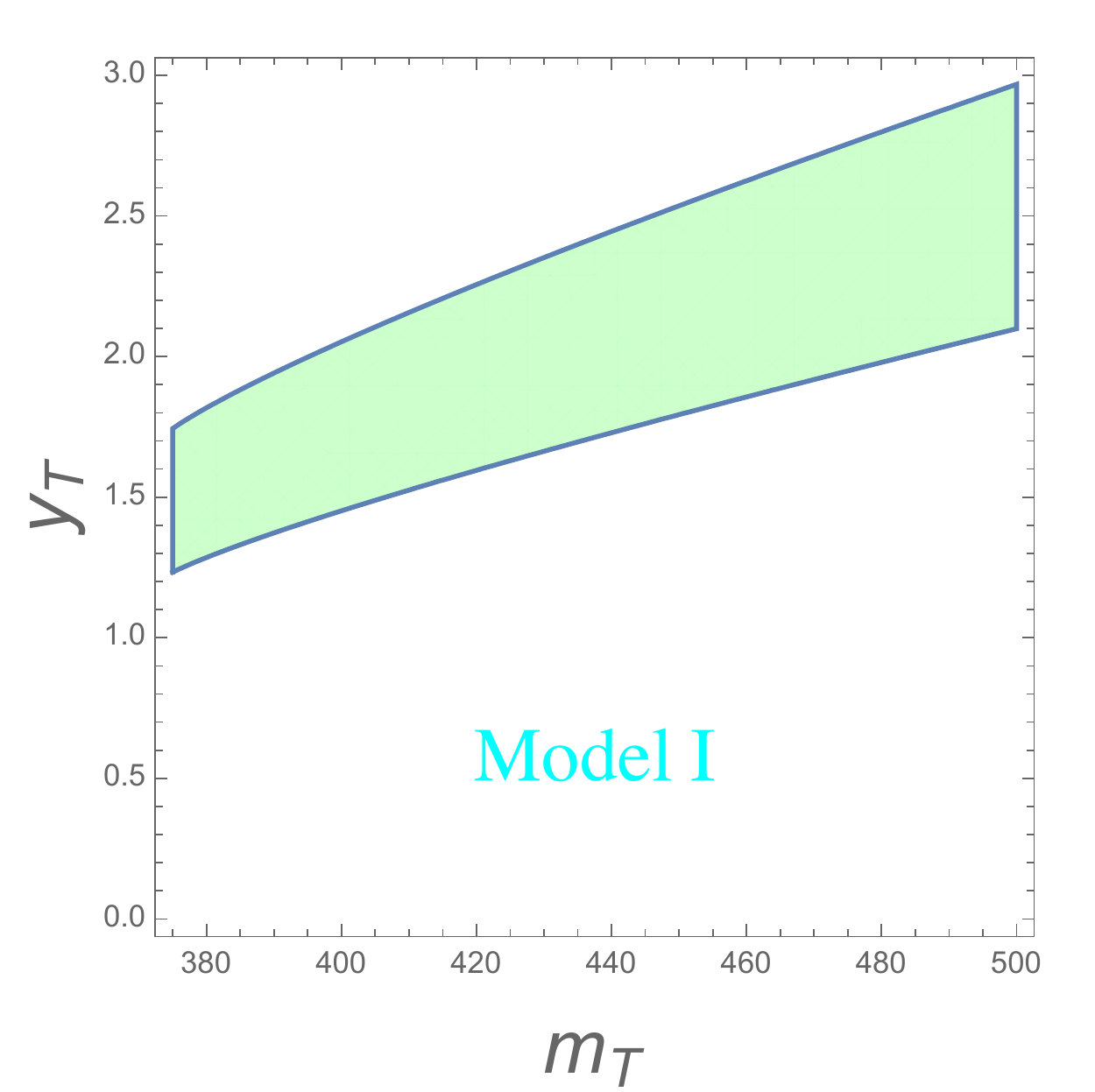}
  \includegraphics[width=0.45\textwidth]{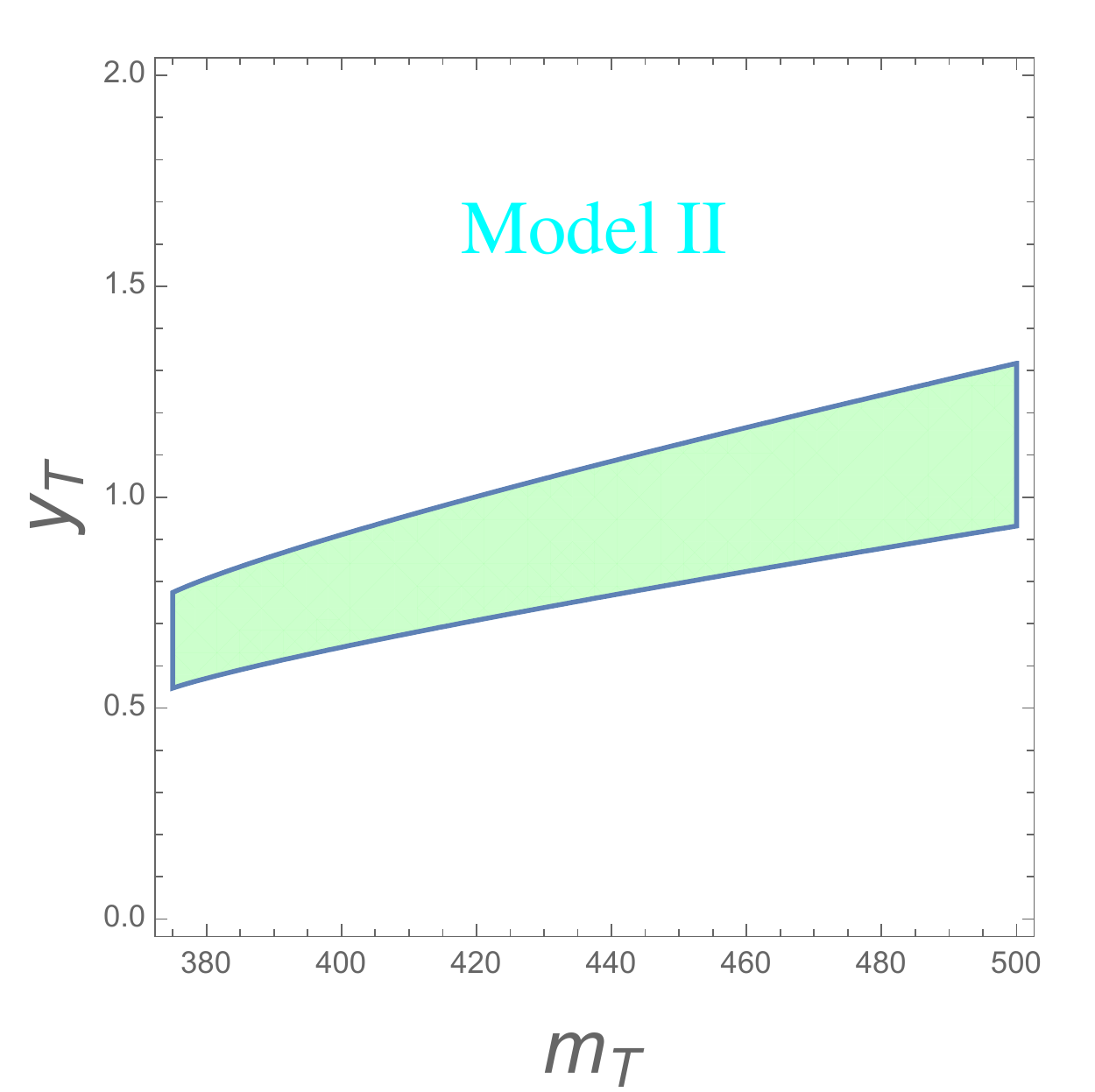}
\caption{\label{region}  The signal favored region $\sigma (gg\to S)\times BR(S\to \gamma \gamma )$ $\in$ (5,10) fb for model I (left panel) and model II (right panel).}
\end{figure}

For the numerical simulation, we work in two alternative scenarios: Model I, which only contain one electroweak singlet VL top quark, and Model II, which has two VL quarks: a electroweak doublet and a singlet.
In Fig.~\ref{region} we show regions in the $y_T-m_T$ plane that is favored by the signal cross section $\sigma (gg\to S) BR(S\to \gamma \gamma )$ $\in$ (5,10) fb for model I (left panel) and model II (right panel).
We have assumed that VL quarks have nearly degenerate masses and the Yukawa couplings with the $S$ are the same for two VL quarks.
Since model II contains three VL tops, the production cross section is greatly enhanced comparing to the case of the model I, thus a smaller Yukawa coupling is able to give rise to a sizable cross section.

\begin{figure}
  \includegraphics[width=0.45\textwidth]{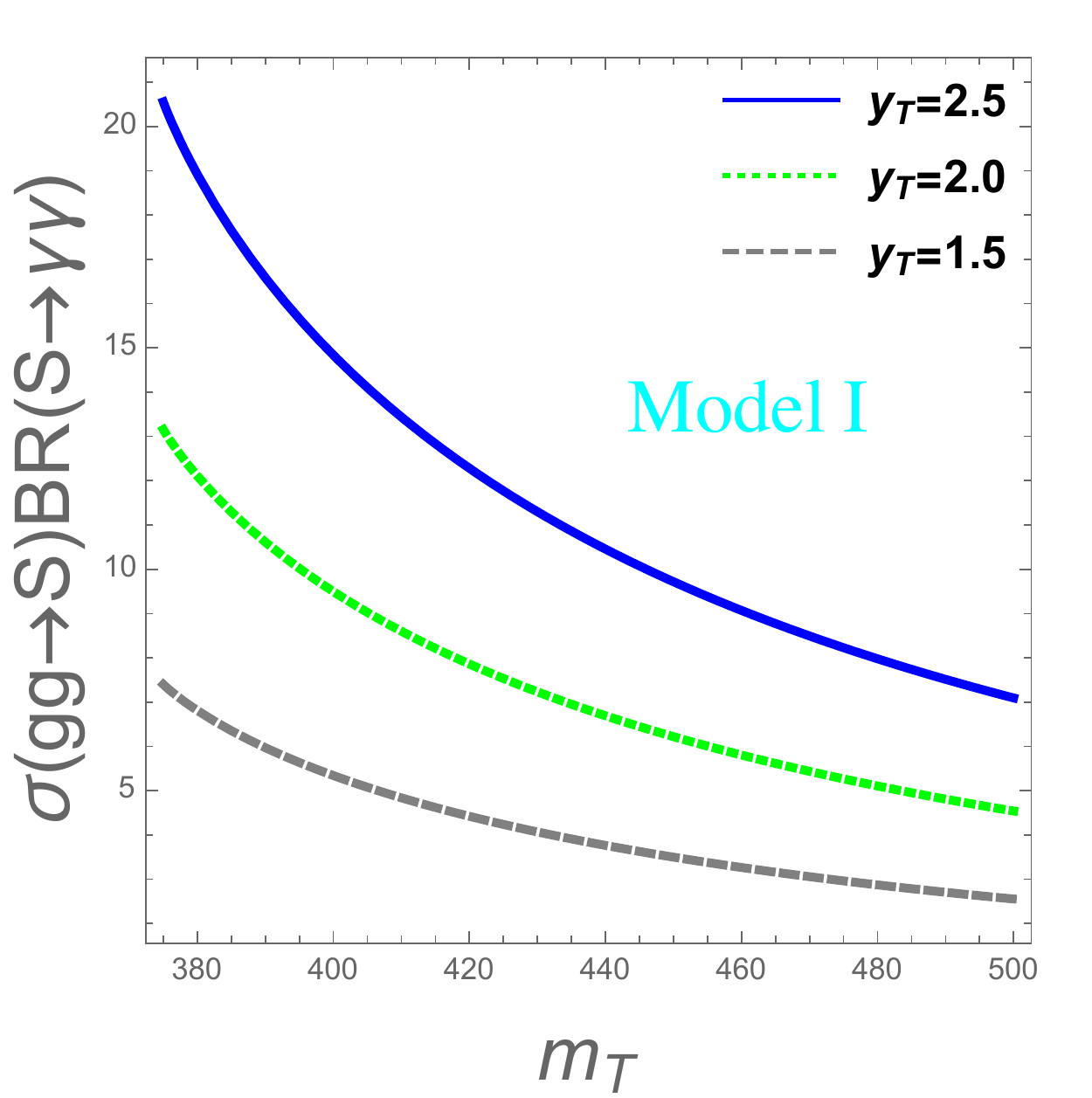}
  \includegraphics[width=0.45\textwidth]{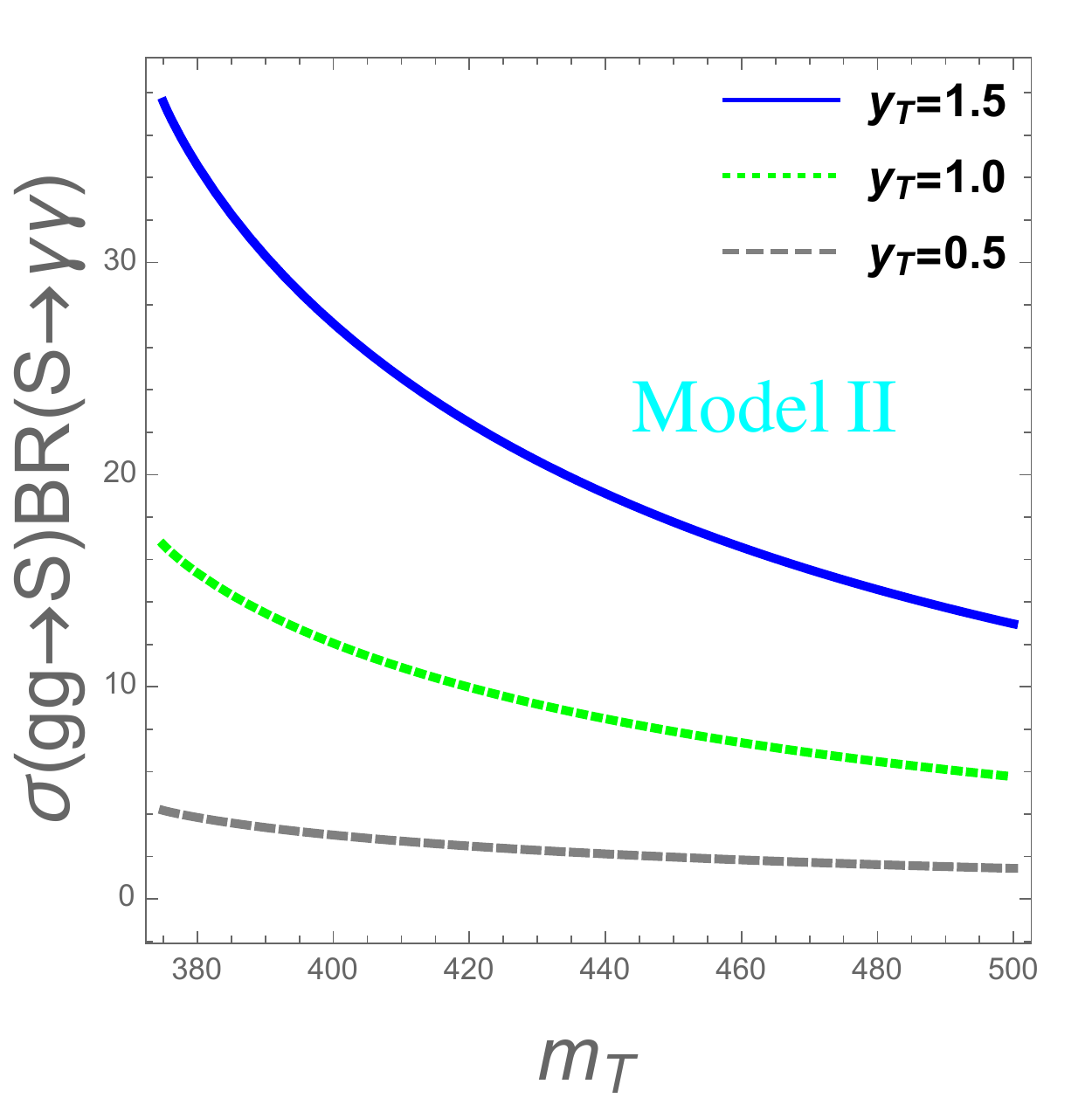}
\caption{\label{cross}   Production cross section times branching diphoton ratio as the function of the VL top mass for model I (left panel) and model II (right panel).}
\end{figure}

%
We show in Fig.~\ref{cross} the cross section $\sigma (gg\to S) {\rm BR}(S\to \gamma \gamma)$ as the function of the VL top(s) mass $m_T$ of model I  ( left panel ) and model II (right panel) respectively.
The solid, dotted and dashed lines correspond to $y_T=2.5, ~2.0,~1.5$ for model I and  $y_T=1.5, ~1.0,~0.5$ for model II.
Apparently one has more accessible  parameter space in model II.
It should be noticed that ${\rm BR} (S\to \gamma \gamma) \sim 0.7\%$, which is only a very small fraction.
The total decay rate of $S$ is  unlikely compatible with the best-fit value of the ATLAS.

\section{Validation on Other Channels}\label{OtherChannel}

\begin{table}[th]
\begin{tabular}{cccccc}
\hline
~~~$\sigma \cdot {\rm Br}({\rm dijet})$~~~ & ~~~$\sigma \cdot {\rm Br}(ZZ)$~~~ & ~~~$\sigma \cdot {\rm Br}(WW)$~~~ & ~~~$\sigma \cdot {\rm Br}(hh)$~~~ & ~~~$\sigma \cdot {\rm Br}(\bar tt)$~~~ & ~~~$\sigma \cdot {\rm Br}(inv.)$~~~ \\
1800 fb~\cite{CMS8diphoton} & 22 fb~\cite{Aad:2015kna} & 38 fb~\cite{Aad:2015agg} & 35 fb~\cite{Aad:2015xja} & 600 fb~\cite{CMS8tt} & 3000 fb~\cite{Aad:2015zva}\\
\hline
\end{tabular}
\caption{The experimental constraints from each decay channel for LHC at $\sqrt{s}=8$ TeV.}
\label{8constraints}
\end{table}

The experimental constraints from 8 TeV LHC are shown in Table \ref{8constraints} for each channel respectively. Due to the nearly $100\%$ branching ratio the $8$ TeV dijet constraint is the most important one, which deserves a closer look.
Numerical calculations are shown in Fig.~\ref{cross}, which shows a loose constraint on our model.
Actually it is a weighted combination of the a looser gluon gluon jet bound $\sigma \cdot {\rm Br}({\rm dijet}) < 2200 \ {\rm fb}$
and a more stringent quark quark jet bound $\sigma \cdot {\rm Br}({\rm dijet}) < 800 \ {\rm fb}$.
We can see that according to Eq.~(\ref{ggtogaga}), the diphoton channel $8$ TeV constraint of $1.32$ fb (to which we respect) corresponds to a dijet cross section of about $200$ fb, one order of magnitude below the bound from the $8$ TeV measurement. For the $13$ TeV  LHC there is no experimental upper bound in the dijet channel.

\begin{figure}
  \includegraphics[width=0.45\textwidth]{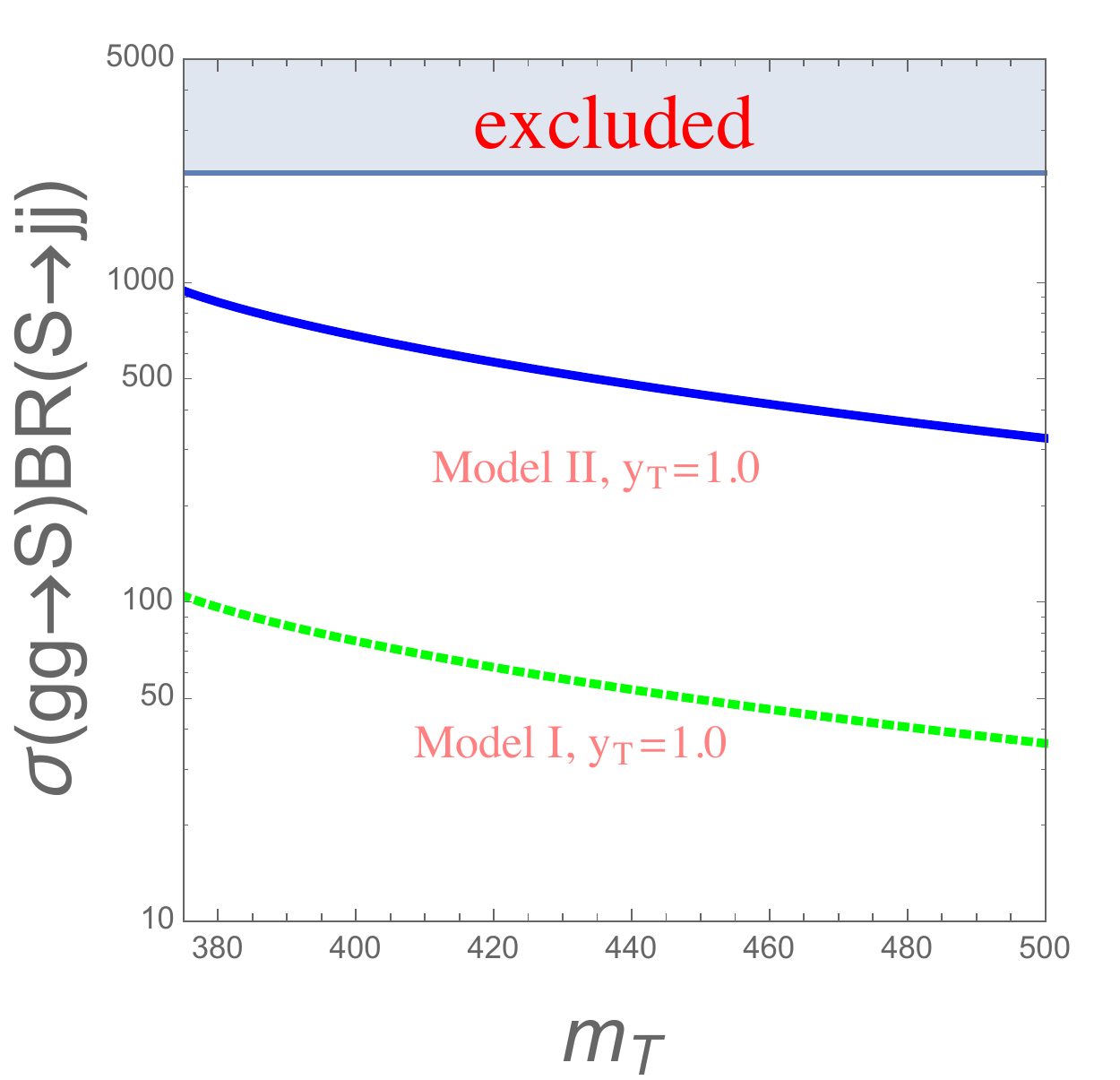}
\caption{\label{dijet} Constraints on the model from the LHC Run-1 in the dijet channel: the cyan region has $\sigma\cdot {\rm BR (dijet)} >2200 ~{\rm fb}$. The solid and dashed line correspond to predictions of model I and model II  respectively with $y_T=1$.    }
\end{figure}

Next we consider constraints from the $WW/ZZ/\bar tt$ channel. In model I the VL top loop also induces a $ZZ$ ($Z\gamma$) decay final state, and in model II the doublet will further induce a $WW$ decay final state, but they are at most the same order compared with the diphoton channel. 
In model I, we estimate the partial widths of $ZZ$ ($Z\gamma$) via the relation
\bea
	\Gamma_{\gamma\gamma} : \Gamma_{Z\gamma} : \Gamma_{Z Z} = 1: 2 \tan^2\theta_W : \tan^4\theta_W 
\simeq 1: 0.6 : 0.09,
\eea
where $\theta_W$ is the weak mixing angle.
For a heavy SM Higgs the tree level induced $WW/ZZ$ channel are actually the dominant decay mode and the Higgs total width is rather broad, that implies our scaled up version of Higgs $S$ may have a large tree level decay width to $WW/ZZ$, if the coupling is not significantly suppressed compared with the SM one. In our model $S$ couples to other SM particles only through the mixing of Eq.~\ref{mixing}, the mixing angle $\varphi$ (see Ref.~\cite{Xiao:2014kba}) satisfies $\tan2\varphi=\frac{\lambda_{HS}v}{\lambda_Su}$, where $u$ is the vacuum expectation value (VEV) of $S$ and $v\simeq246$ GeV is the SM Higgs VEV. The mixing could be small due to a parametrically small $\lambda_{HS}$, in that limit the linear combination of the 750 GeV S is dominated by the heavier interaction eigenstate.
The tree-level partial width of the $S$ can be expressed by a scaling from the prototype of the SM heavy Higgs decay
\bea
\Gamma(S\to WW/ZZ/t\bar t)=\Gamma(h_{m_h=750\text{GeV}}\to WW/ZZ/\bar tt) \sin^2\varphi.
\eea
The partial decay width of a $750~{\rm GeV}$ SM-like Higgs boson is given in Ref.~\cite{Heinemeyer:2013tqa}. Taking the $ZZ$ channel for example, the requirement of $\sigma \cdot {\rm Br}(ZZ)<22$ fb corresponds to the tightest constraint on the mixing angle $\varphi$ between the Higgs boson and the new scalar:
\bea
	\sin\varphi \simeq \frac{\lambda_{HS}v}{2\lambda_Su} < 6.7 \times 10^{-3}.
\eea
Similarly for the $WW$ channel,  the upper limit on the mixing angle is about $8.5 \times 10^{-3}$. For the hh channel the upper limit is $1.1 \times 10^{-2}$.
Furthermore, the invisible decay channel could increase the total width of the 750 GeV scalar.
Given the VL top mass around 400 $\sim$ 500 GeV and the stealth region defined in Eq.~(\ref{eq:stealthmass}), the dark matter mass is around $220 $ $\sim$ 490 GeV depending on their mass difference. 
Although the diphoton branching ratio will decrease if the invisible decay channel is open, if we increase the production cross section by adding more VL quarks, the total decay width could be greatly enhanced.
Here we just simply neglect this possibility, by assuming the dark matter mass is heavier than half of the scalar mass or coupling of $S$ to dark matter is negligible.

\section{conclusion}

To explain the diphoton excess reported by the ATLAS and CMS at  $\sqrt{s}=$ 13 TeV recently,
we investigate the possible explanation based on the vectorlike quark framework.
To fit the observed diphoton rate, the vectorlike quark needs to be not so heavy, otherwise it leads to a too large Yukawa coupling of the 750 GeV scalar to the vectorlike quark.
However, the LHC searches push the vectorlike quark mass heavy.
We investigated a special vectorlike quark: the stealth top, in which the vectorlike top is $Z_2$-odd and has degenerate mass with the dark matter.
In this setup, the particle content could be very minimal: a singlet scalar and stealth top.
We also discussed the case when additional vectorlike doublet is added.
We examined the constraints on the stealth top from the LHC and LEP searches, and found the constraints are quite loose.
We found that a stealth top with mass around $400 \sim 500$ GeV  and its coupling to the scalar is around $0.5 \sim 1.5$, could fit the diphoton signal well.
We commented that our economic setup can be extended to more complicated UV models.

To validate the diphoton signature, we examined the current LHC constraints on other channels such as $WW$, $ZZ$ and $jj$, and found that we manage to enhance the diphoton signature while keeping other channels far from the current constraints.
We expect that the future Run-2 LHC could cross check the signature from other decay channels.

\begin{acknowledgments}
This work of WC and JHY was supported in part by DOE Grant DE-SC0011095. RH was supported in part by the World Premier International Research Center Initiative, Ministry of Education, Culture, Sports, Science and Technology, Japan.
\end{acknowledgments}

\end{document}